\documentclass[a4paper,twocolumn,11pt]{quantumarticle}
\pdfoutput=1
\usepackage[a4paper, total={6in, 8in}]{geometry}
\usepackage{amsmath,amsthm,amssymb,amsfonts,wrapfig,graphicx,indentfirst,tikz,verbatim}
\usepackage{physics}
\usepackage[T1,T2A]{fontenc}
\usepackage[utf8]{inputenc}
\usepackage[english]{babel}
\usepackage[all]{xy}
\usepackage{hyperref}
\usetikzlibrary[angles,quotes]

\newcommand{\sam}{\mathcal S}

\title{A state chaining-based objective collapse model}
\author{Roman V. Li}
\email{roman.lithium@gmail.com}
\orcid{0009-0004-0146-8665}
\affiliation{V. V. Voevodsky Institute of Chemical Kinetics and Combustion of the Siberian Branch of the RAS, 	Institutskaya Street 3, Novosibirsk, 630090, Russia}

\begin{document}

\maketitle

\begin{abstract}
    The quantum-to-classical transition hinges on the nature of wavefunction collapse, which remains a central controversy in foundational physics. Objective collapse theories aim to modify quantum mechanics by introducing a physical, non-subjective mechanism for irreversible events, but existing models face significant conceptual and empirical challenges. Here, we propose a novel collapse mechanism based on a specific form of quantum correlation termed ''chaining'', formalized within a new diagrammatic framework (quantum illustrations, or qils). This approach does not rely on system size or environmental complexity, but on the probabilistic occurrence of a collapse event with a fixed, universal probability $1/\Sigma$ per chaining step. We demonstrate that this model naturally explains the emergence of classicality in paradigmatic scenarios (measurement devices, Schrödinger’s cat, spontaneous decay) and makes testable predictions for interference experiments. The theory is shown to be consistent with existing data from delayed-choice quantum eraser and matter-wave interference experiments, yielding an estimate for the fundamental constant $\Sigma \geq 1.5$. By providing a unified, parameter‑sparse mechanism for objective collapse, this work bridges quantum and classical descriptions and has implications for the interpretation of quantum experiments, the design of quantum computers/sensors, and the understanding of decoherence in complex systems.
\end{abstract}

\section{Introduction}

The measurement problem is a well-known topic for discussion in quantum mechanics. Starting from Einstein, Podolsky and Rosen \cite{Einstein1935}, and then Schrödinger \cite{Schrodinger1935}, physicists have been continuously researching the boundaries of the quantum domain. The Schrödinger's renowned argument (''Schrodinger's cat'') provided us with a method to expand quantum superposition onto a macroscale system through connecting states of larger and larger systems with an unstable atom. Then, since just one choice (a state of an atom) chooses between two macroscopic scenarios, the superposition of the atom's states creates a superposition of these scenarios.

Since Schrodinger's time, multiple interpretations arose, having different opinions of how to deal with this conundrum. The Copenhagen interpretation states that the superposition will remain until it meets with a measurement device \cite{Baggott:2011zz}; however, it gives no criteria on how to distinguish that device from a simple bunch of atoms, which is, obviously, able to inherit the superposition from the system considered.

On the other hand, one may consider the device to be a quantum system. Then, that device, being in superposition itself, may be observed by the next device and so on. It makes no difference in the predictions of quantum theory where along this chain of causal effects the superposition collapses \cite{Wheeler2017}. The ''consciousness causes collapse'' argument states that this superposition eventually meets a consciousness and collapses since a consciousness cannot be multiple \cite{OMNS2020}. Thus, the consciousness is necessary for a wavefunction to collapse.

The Everett's interpretation, however, states the opposite. The consciousness \textit{does} inherit the superposition; however, it's impossible to access every other scenario in the superposition set from the chosen one, since quantum decoherence destroys every quantum interactions between them, converting the entangled state into a mixed one \cite{Zurek2003}\cite{zurek2003decoherencetransitionquantumclassical}.

However, these and many others interpretations don't answer the question \textit{how} one could predict the experimental outcome for a given setup. The Born rule arises ''somewhere in between'' the quantum object and the observer without any specification \textit{where} that trigger boundary lies. The ''consciousness causes collapse'' postulate seemingly presents such a specification; however, without any conventional rule allowing to distinguish conscious and unconscious systems (see Chalmers \cite{Chalmers1996-CHATCM-18} and Kirk \cite{Kirk2005-KIRZAC-2}) this interpretation makes no use. Also, there are several arguments against it, referencing to materialistic and heuristic arguments \cite{Penrose1989}\cite{schreiber1995livesschroedingerscat}.

However, the attempts are made to add wavefunction collapse as a part of an objective reality. In these \textit{objective collapse theories} superposition is destroyed naturally by a process of time evolution, that differs from a unitary Schrodinger's law \cite{Okon_2014}. In these models the quantum/classical threshold emerges automatically from some modifications added to a unitary evolution. A well-known theory in this field is the \textit{continuous spontaneous localization} (CSL) model \cite{Pearle1989}\cite{Ghirardi1990}, where collapse takes place in a position basis. Being negligible at the microscopic level, its effect becomes stronger at the scale of macroobjects, thus leading to their emergent classicality.

A special case is a hypothesis made by Diosi \cite{Disi1989} and Penrose \cite{Penrose1996} (DP-model), where gravity plays a role of that one classical entity needed for a Copenhagen interpretation to work. All these researches form a special category since they attempt not to \textit{reinterpret} quantum mechanics, but to expand its laws and make a new \textit{theory}. The key term in these theories is an \textit{event} - a non-unitary irreversible change of a quantum state (such as acts of measurement).

Presented in this article is a formalism of \textit{qils} (''\textbf{q}uantum \textbf{il}lustrations'') that describes states of quantum systems from the point of view of their \textit{degrees of freedom}. This formalism allows us to account the only aspects of quantum systems that matter in the consideration of events, and then propose a new model of objective quantum events based on the concept of \textit{quantum chaining} of subsystems' degrees of freedom. This model predicts a universal collapse not depending on the system's size parametrized by the universal constant $\Sigma$. Then we illustrate, how the formalism of qils describes some conventional experimental setups and how events arise in measurement devices and processes such as spontaneous decay. Finally, we derive a collapse effect, find its presence in different experimental data and show that they are in good correspondence with our model.

\section{Motivation}

In this article a solution to a measurement problem through considering \textit{quantum events} is proposed. The events should have the following properties \cite{Frantsuzov2020}:
\begin{itemize}
    \item An event is not necessarily a measurement;
    \item A quantum measurement is a specially constructed event;
    \item Events are irreversible;
    \item An event is an objective part of physical reality;
    \item Events are discrete and are classical-like (e. g. they can't be in a superposition of their possible outcomes);
    \item Some events are non-local;
    \item Events are described by a new mechanism, that isn't included in a law of unitary evolution.
\end{itemize}

The law governing the events should be constructed to be in agreement with all known phenomena, such as
\begin{enumerate}
    \item A measurement device guarantees a state collapse (a special case of events);
    \item Macroscopic objects don't have their observable (\textit{external}) parameters in a superposition;
    \item Different microscopic objects behave alike in interference experiments regardless of their internal structure (see Section \ref{ch:experimental});
    \item There're macroscopic objects that don't cause collapse, such as mirrors.
\end{enumerate}

The considerations following are based on the observation than any measurement device, being considered as a quantum object, splits into a superposition of states corresponding to a set of states of the measured object. This set contains much less elements than could be represented by a device. For example, if a measurement device with a pointer measures a spin projection of an electron, there are only two states for the pointer, instead of every combination of the pointer's particles being in a state 'up' or 'down' in an arbitrary combination. This observation leads us to a notion that events are somehow related to a degeneracy of a system's state set.

However, the third point of the list above hints on the fact, that events can't be directly connected on internal complexity of a system. Instead, the event mechanism must be based only on the external parameters, such as a net momentum of a particle or a total spin of a molecule.

Thus, the Section \ref{sec:theses} introduces an event mechanism, that satisfies these requirements. Instead of relying on the system's size directly, the mechanism considers probabilities of events on every step of propagation of a superposition, if that propagation leads to a state set degeneracy. For example, the Schrodinger's cat causes collapse not because of its macroscopic size, but because it has a lot of subsystems with only two shared, correlated states, and these subsystems are included into a superposition by lots of steps. Indeed, if an object is macroscopic, it usually requires lots of steps to be involved into a superposition completely; however, the resulting state set will contain much less elements than a total possible state set of an object. Since the greater size requires the greater number of acts of propagation of a degenerated superposition, macroscopic objects could trigger an event with much greater probability, if an event has a probability to occure on every step of that propagation. And, of course, this doesn't rely on a size of a system directly and allows the systems of different sizes behave the same in interference experiments (there's only one step of superposition propagation - the beam splitter).

\section{Qil-formalism}

In order to expose that one criterion that leads to quantum events, we will now develop a diagram technique that will be complementary to usual notations (such as a bra-ket notation or density matrices). That \textit{qil-formalism} will be focused on description of number of states corresponding to different parametrizations of quantum state spaces.

A \textit{qil} (''\textbf{q}uantum \textbf{il}lustration'') is a symbol consisting of the body (a qil's name in brackets) and indices:

\begin{eqnarray*}
    (q)^{\alpha_1...\alpha_n}_{\beta_1...\beta_m}
\end{eqnarray*}

Every index corresponds to a some basis in a state space, corresponding to a certain parameter of a system, to which we'll be referring as \textit{degree of freedom} (DoF). This basis contains either a single permitted state with a certain DoF's value (upper index, the corresponding DoF is \textit{defined}) or a set of different states with different values of the corresponding DoF (lower index, the DoF is \textit{undefined}). For example, the state of a system $q$ with a defined momentum $p$ has a state set $\sam_p(q)$ (corresponding to a $p$-parametrization) with a singular permitted state $\ket{p}$; meanwhile, the  $\sam_x(q)$ (a set of possible coordinates) of that system has a continuum of different equally probable states $\ket{x}$. Thus, the corresponding qil then has the following form:

\[(q)_x^p\]

with a state set

\[\quad |\sam_p(q)| = |\{\ket p\}| = 1,\]
\[|\sam_x(q)| = \left|\left\{\ket x\ \bigg|\ x \in \mathbb R\right\}\right| > 1.\]

An important example that will be frequently used in the present article is a qil for a \textit{quantum mode} with an occupation number $n$. For example, if the particle could get into a left or a right slit in a double-slit experiment, the right-slit mode of a trajectory would have a following qil:

\[(r)_n\]

with a state set

\[\sam_n(r) = \left\{0, 1\right\}\]

Here the right-slit mode could have either $0$-value (if the particle goes to the left) or $1$-value (if the particle goes to the right).

We'll also use the qil $(q)_\phi$ with a wavefunction phase $\phi$ as a parameter. It should be considered not as a phase itself (since it's unobservable), but as a phase shift between different parts of a total wavefunction.

The next step in developing the formalism is the introduction of special types of qils - \textit{brackets}, that don't exist themselves, but merge other qils instead. Every bracket corresponds to a certain operation on the state sets involved. The square bracket is called an \textit{entanglement} and constrains the subqils' degrees of freedom of the same type with some rule:

\begin{widetext}

\begin{eqnarray}
    (0)_p = \left.\begin{gathered}
        (1)_p\\(2)_p\\(3)_p
    \end{gathered}\right]_p \quad \longleftrightarrow \quad f(p_1, p_2, p_3) = p_0
\end{eqnarray}

The state set of the systems $[1, 2, 3]$ combined with an entanglement bracket is then

\begin{eqnarray}
    \sam_p[1, 2, 3] = \left\{[p_1, p_2, p_3]\ \bigg|\ f(p_1, p_2, p_3) = p_0\ \text{for}\ p_i\in \sam_p(i)\ \forall i \in \{0, 1, 2, 3\}\right\}
\end{eqnarray}

Two particles after a collision become entangled along their momentum since it's conserved:

\begin{eqnarray}
    \begin{gathered}
        (1)^p\\(2)^p
    \end{gathered} \rightarrow \left.\begin{gathered}
        (1)_p\\(2)_p
    \end{gathered}\right]^p
\end{eqnarray}

Also the entanglement could occur in a case of splitting the system. Consider a $\pi$-meson with a spin $0$ decaying into two photons:

\begin{eqnarray}
    (\pi)^s \rightarrow \left.\begin{gathered}
        (\gamma)_s\\(\gamma)_s
    \end{gathered}\right]^s
\end{eqnarray}

Note that qil-diagrams contain no information about coefficients before terms in exact state expressions. For example, a particle with an undefined vertical spin component has a following qil-diagram regardless of coefficients:

    \begin{eqnarray}
     (e)^n_{s_z} = \left.\begin{gathered}
        (\uparrow)_n^{s_z}\\ (\downarrow)_n^{s_z}
    \end{gathered}\right]^n\quad \longleftrightarrow \quad \ket{1}_e = \alpha\ket{1}_\uparrow\ket{0}_\downarrow + \beta\ket{0}_\uparrow\ket{1}_\downarrow \quad \forall \alpha, \beta.
    \end{eqnarray}
\end{widetext}

where $\ket{n}_\uparrow$ and $\ket{n}_\downarrow$ are a spin-up and a spin-down modes with an occupation number $n$. Here the two modes (spin-up and spin-down) are entangled by their occupation number since the total $n$ is defined and equals $1$.

The natural evolution could lead to an entanglement with two different ways. The first one creates an entanglement with a defined external index

\begin{eqnarray}
    \begin{gathered}
        (1)^p\\(2)^p
    \end{gathered}\rightarrow\left.\begin{gathered}
        (1)_p\\(2)_p
    \end{gathered}\right]^p
\end{eqnarray}

The other way creates an undefined entanglement

\begin{eqnarray}
    \begin{gathered}
        (1)_p\\(2)^p
    \end{gathered}\rightarrow\left.\begin{gathered}
        (1)_p\\(2)_p
    \end{gathered}\right]_p\qquad\text{or}\qquad\begin{gathered}
        (1)_p\\(2)_p
    \end{gathered}\rightarrow\left.\begin{gathered}
        (1)_p\\(2)_p
    \end{gathered}\right]_p
\end{eqnarray}

These processes correspond to a set of systems with an interaction Hamiltonian $\hat H_I(\hat p_1,\hat p_2)$, whose effect increases the total number of states.

On the other hand, the curly bracket (a \textit{chaining}) identifies the corresponding DoFs of the included qils, making them one and the same DoF. It represents a situation where multiple subsystems are forced to share a single, common value of a given parameter:

\begin{eqnarray}
    (0)_p = \left.\begin{gathered}
        (1)_p\\(2)_p\\(3)_p
    \end{gathered}\right\}_p \quad \longleftrightarrow \quad p_i = p_0 \ \ \forall i \in \{1, 2, 3\}.
\end{eqnarray}

The chaining bracket identifies all state sets combined by them:

\begin{eqnarray}
    \sam_p(0) = \sam_p(i)\quad \forall i\in\left\{1, 2, 3\right\}.
\end{eqnarray}

The bracket itself doesn't correspond to any given quantum state expression. Instead, it appears as a part of a description of states like this:

\begin{widetext}
    \begin{eqnarray}
    (\psi)^n = \left.\begin{gathered}
        \left.\begin{gathered}
            (a)_n\\ (b)_n
        \end{gathered}\right\}_n\\ \left.\begin{gathered}
            (c)_n\\ (d)_n
        \end{gathered}\right\}_n
    \end{gathered}\right]^n\quad \longleftrightarrow \quad \ket{1}_\psi = \alpha\ket{1}_a\ket{1}_b\ket{0}_c\ket{0}_d + \beta\ket{0}_a\ket{0}_b\ket{1}_c\ket{1}_d
    \end{eqnarray}
\end{widetext}

Such qils, for example, correspond to compound particles in a beam splitter. Consider the following example (keeping in mind that both components of a particle must go to the same side):

\begin{eqnarray}
\label{eq:ex_bs}
    (\pi)^n = \begin{gathered}
        (u)^n\\(\bar u)^n
    \end{gathered}\rightarrow \left.\begin{gathered}
        \left.\begin{gathered}
            (u|1)_n\\(\bar u|1)_n
        \end{gathered}\right\}_n\\\left.\begin{gathered}
            (u|2)_n\\(\bar u|2)_n
        \end{gathered}\right\}_n
    \end{gathered}\right]^n
\end{eqnarray}

where $(u|1)$ and $(u|2)$ are the qils of a first and a second trajectory after the beam splitter.

Since a chaining identifies all the state spaces considered and the natural evolution forbids the number of states to be reduced, the chaining in a pair interaction could occur only if at least one of the interacting systems has a defined DoF:

\begin{eqnarray}
    \begin{gathered}
        (1)^p\\(2)^p
    \end{gathered}\rightarrow\left.\begin{gathered}
        (1)_p\\(2)_p
    \end{gathered}\right\}_p\qquad\text{or}\qquad\begin{gathered}
        (1)^p\\(2)_p
    \end{gathered}\rightarrow\left.\begin{gathered}
        (1)_p\\(2)_p
    \end{gathered}\right\}_p
\end{eqnarray}

Also, a chaining could be a result of defining a new DoF, as it was in a beam-splitter example \eqref{eq:ex_bs}:

\begin{eqnarray}
    \begin{gathered}
        (a)\\(b)
    \end{gathered} \rightarrow \left.\begin{gathered}
        \left.\begin{gathered}(a|1)_n\\(b|1)_n
    \end{gathered}
        \right\}_n\\\dots\\\left.\begin{gathered}
            (a|k)_n\\(b|k)_n
        \end{gathered}\right\}_n
    \end{gathered}\right]^n
\end{eqnarray}

where $(a|k)_n$ is a qil of a system $a$ in a state $k$ with an occupation number $n$.

The definition of the brackets allows us to state the \textit{reduction rules}:

\begin{equation*}
\label{eq:reduction}
\left.\begin{gathered}
(1)^{p}\\ (2)_{p}\\ (3)_{p}
\end{gathered}\right]_p = \begin{gathered}
(1)^p\\ \left.\begin{gathered}
(2)_{p}\\ (3)_{p}
\end{gathered}\right]_p
\end{gathered}
\end{equation*}

\begin{equation*}
\left.\begin{gathered}
(1)^{p}\\ (2)_{p}\\ (3)_{p}
\end{gathered}\right]^p = \begin{gathered}
(1)^p\\ \left.\begin{gathered}
(2)_{p}\\ (3)_{p}
\end{gathered}\right]^p
\end{gathered}
\end{equation*}

\begin{equation*}
\left.\begin{gathered}
(1)^{n}\\ (2)_{n}\\ (3)_{n}
\end{gathered}\right\}_n = \begin{gathered}
(1)^{n}\\ (2)^{n}\\ (3)^{n}
\end{gathered}
\end{equation*}

\begin{equation}
\left.\begin{gathered}
(1)_{n}\\ \left.\begin{gathered}
(2)_{n}\\ (3)_{n}
\end{gathered}\right\}_{n}
\end{gathered}\right\}_{n} = \left.\begin{gathered}
(1)_{n}\\ (2)_{n}\\ (3)_{n}
\end{gathered}\right\}_{n}
\end{equation}

Also for the event theory to be constructed let's define an \textit{event} as an internal transition of any undefined degree of freedom into its defined state:

\begin{eqnarray}
    (\psi)_{p\uparrow} = (\psi)^p.
\end{eqnarray}

The DoF obtains one of its values that was in its state set before the event. The probability of the specific value is determined by the Born rule based on a wavefunction corresponding to the qil. Such event is, in fact, the moment when ''God plays dice''.

After the event, the reduction rules \eqref{eq:reduction} must be used to simplify the obtained diagram. Note that quantum nature of defined DoFs causes their canonical conjugates to become totally undefined:

\begin{eqnarray}
    (\psi)_{p\uparrow}^x = (\psi)^p_x.
\end{eqnarray}

This also causes the qil to leave all the brackets that were merging it through that DoF.

In the Table \ref{t:qil_examples} some examples of quantum processes and their qils are listed.

\begin{table*}
\centering
 \begin{tabular}{|p{8cm} | p{5cm}|} 
 \hline
 Entity & Qils \\ [0.5ex] 
 \hline\hline 
 A collision of two particles with defined initial momenta with the total momentum conserved. & \begin{center} $\begin{gathered}
     \begin{gathered}
     (1)^p\\(2)^p
    \end{gathered}\rightarrow\left.\begin{gathered}
     (1)_p\\(2)_p
    \end{gathered}\right]^p\\\,
 \end{gathered}$
 \end{center} \\
 \hline 
 A photon before and after interaction with a translucent mirror. The modes $(t)$ (transition) and $(r)$ (reflection) are entangled by their occupance number $n$. & \begin{center}
     $(\gamma)^n\rightarrow \left.\begin{gathered}
     (t)_n\\(r)_n
    \end{gathered}\right]^n$
 \end{center} \\
 \hline
 A $\pi^0$-meson with spin $s = 0$ decaying into two photons with some probability. There are two modes (remaining $\pi^0$ and decay products) which are constrained (entangled) by their occupation number $n$. Two photons are also chained since they either exist simultaneously or both don't exist. & \begin{center}
     $(\pi^0)^{sn}\rightarrow \left.\begin{gathered}
         (\pi^0)_n^s\\ \left.\left.\begin{gathered}
     (\gamma_1)_{sn}\\(\gamma_2)_{sn}
    \end{gathered}\right]^s\right\}_n
     \end{gathered}\right]^n$
 \end{center} \\
 \hline
 \end{tabular}
 \caption{the examples of qil-diagrams.}
\label{t:qil_examples}
\end{table*}

\section{Theses}
\label{sec:theses}

The model proposed will be based on the following statements (theses):
\begin{enumerate}
    \item \textbf{All qils are equivalent}. This thesis states that a qil as an object is described \textit{only} by its external degrees of freedom, regardless of any structure. This does not exclude interactions with subqils (like interactions of an electron in an atom with an external photon); it just states that such processes must not be considered as interactions with a qil itself.
    \item \textbf{Every qil interaction occurs through a set of degrees of freedom}. Interactions of systems lead to appearance of one or several brackets and form new qils out of interactants' qils. The interaction's DoFs form a new qil's external DoFs. A type of the brackets formed is dictated by the interaction itself: thus, an entanglement doesn't change the total count of degrees of freedom (reducing the number of internal DoFs by $1$ and creating new external one) and a chaining replaces any number of degrees of freedom with one external (''a chaining of DoFs'').
    \item \textbf{Chainings trigger events}. A process leading to a simultaneous appearance of one or several new chainings along a single degree of freedom has an option to form them with an upper index (to trigger \textit{an event} and thus a quantum state collapse). This option is governed by some universal rule.
\end{enumerate}

The first thesis may be considered as an analogue of ''scale-invariance'' since it allows us to treat equally systems of any size and complexity. It states that a composite qil is, in fact, \textit{a bracket itself} but not a set of subqils. The motivation for this thesis was the fact that photons, neutrons, and atoms behave alike in the same setup from the event theory's point of view (e.g. an atom doesn't lose coherence faster than a neutron in interference experiments): a property that doesn't emerge clearly without the first thesis.

The second thesis may seem quite tautological since the only properties qils have are their degrees of freedom. However, it postulates a basis for their interactions, limiting them into a set of two fundamental brackets - a chaining and an entanglement. It could be easily shown that any complicated correlations between interactants' degrees of freedom can be expanded in terms of these two brackets.

Finally, the third thesis is what makes the theory work. It connects events to chainings, which are stating, in fact, that after the interaction the total number of degrees of freedom the collective of qils gets is less than it needs. The third thesis distinguishes the measurement devices (e.g. a device that lights a whole screen when the particle hits it; it chains the degrees of freedom of all the pixels to a one particle) from any other set of quantum systems (such as a translucent mirror, which has its net momentum as the only degree of freedom that interacts with a photon in the end, and it's entangled, not chained). It also leaves us room to maneuver since it doesn't state the exact rule. However, in this paper we suggest the following principle for the third thesis:

\textbf{''Chinese Whispers rule''}. An appearance of one or several new chainings has a $\frac{1}{\Sigma}$ probability to trigger an event. The $\Sigma$ is a fundamental constant of the theory and doesn't depend on a process or qil properties.

The name of the rule reflects the fact that all events are identical and occur as a chaining translates (''whispers'') one qil's degree of freedom to the other qil. This constant, universal probability per chaining $1/\Sigma$ is the minimal postulate consistent with Thesis 1 (equivalence of all qils). Since a qil is defined only by its external degrees of freedom, the mechanism triggering an event cannot depend on the internal structure, energy, or any other property of the system besides the mere fact of a chaining occurring.

The following analysis in this paper will be done with Chinese Whispers (''ChiWhi'') rule being put as default.

\section{Examples}

\subsection{Delayed choice quantum eraser (qils of modes in superposition)}

\begin{figure*}
    \centering
    \includegraphics[width=0.8\linewidth]{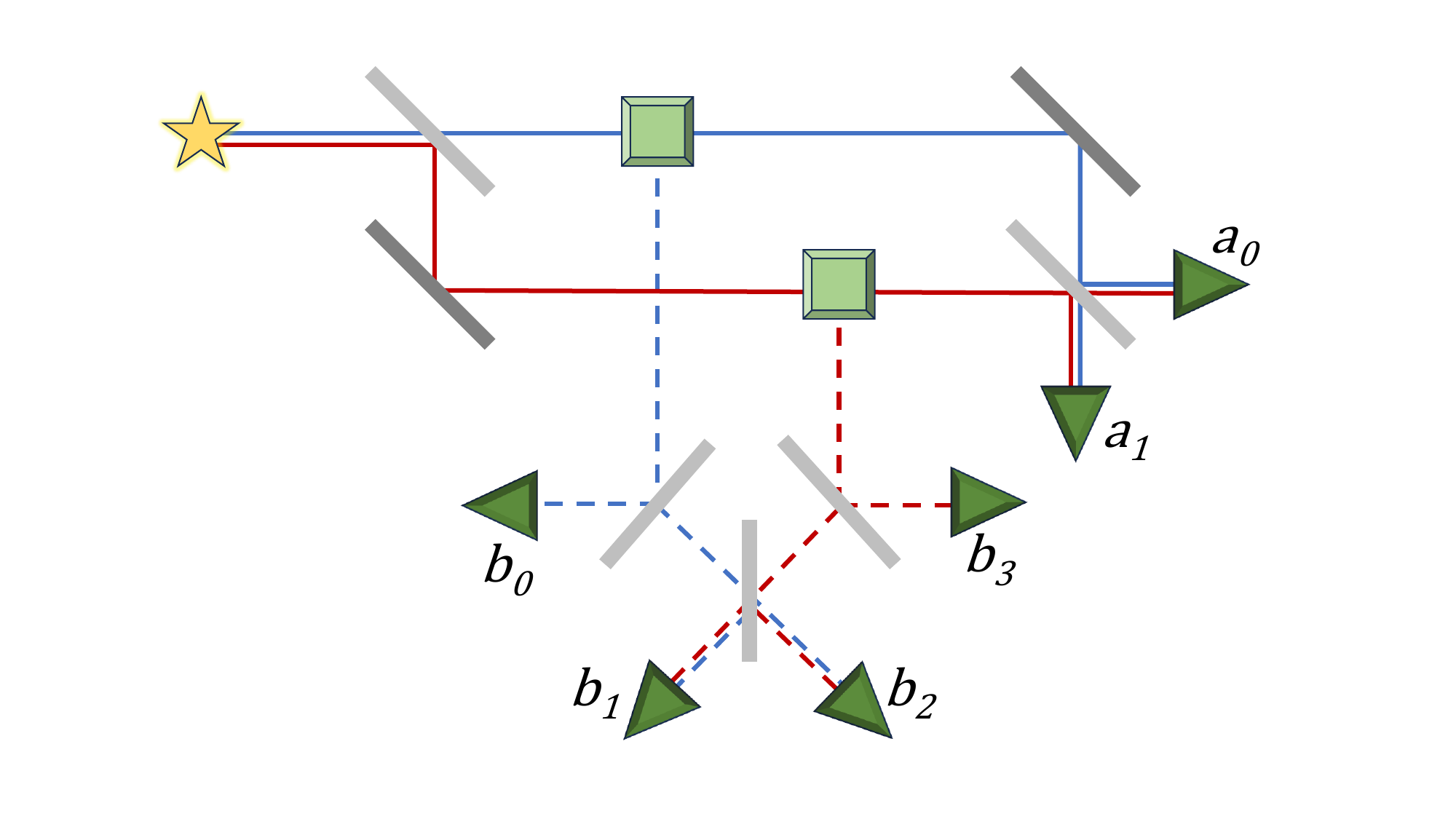}
    \caption{A scheme of a delayed-choice quantum eraser experiment based on a Mach - Zehnder interferometer. The star is a single-photon sourse; the thin rectangles depict mirrors; the cubes represent pieces of nonlinear crystals; and the triangles are detectors. Different trajectories are depicted by different lines for them to be more distinguishable.}
    \label{fig:eraser}
\end{figure*}

The following example is intended to help to get used to the diagram technique. Depicted on the Fig. \ref{fig:eraser} is a scheme of a variant of a renowned \textit{delayed-choice quantum eraser} experiment \cite{dcqe}. However, instead of using a double-slit splitter, this scheme is based on a Mach - Zehnder interferometer \cite{zehnder}. Without any modifications, this interferometer has a $1:0$ ratio between the detectors $a_0$ and $a_1$. This is considered to be a pure interference pattern.

The scheme considered introduces a following modification. There are two nonlinear crystals (the cubes on the scheme) that convert a single photon into an entangled pair. One photon of the pair becomes a \textit{signal photon} and continues its path through the interferometer, and another photon (an \textit{idler photon}) goes into the lower part of the scheme. There both idlers visit a system of mirrors and finally reach the $b$-detectors.

One must notice the following:
\begin{itemize}
    \item If the $b_0$ detector registers a photon, the photon must originate from the upper path. Thus, the wavefunction collapses, and the $a_0:a_1$ ratio becomes $0.5:0.5$;
    \item If the $b_1$ clicks, the idlers are in phase. That means the signal photons are in phase too, and $a_0:a_1 = 1:0$;
    \item The $b_2$ click causes $a_0:a_1 = 0:1$;
    \item And, finally, $b_3$ corresponds to $a_0:a_1 = 0.5:0.5$.
\end{itemize}

When started, the experiment counts coincidences between $a$'s and $b$'s. This allows to divide the $a$-counts into four groups that correspond to the four $b$-detectors and extract the expected patterns.

The conventional interpretation of this experiment assumes that the $b$-detectors define the interference pattern on the $a$-detectors even if they do their measurement later. Let's construct a diagram of the experiment and show that there is no retrocausality needed to explain the statistics:

\begin{widetext}
\begin{eqnarray}
    (0)^{n\phi} 
\overset{1}{\to} \left. \begin{gathered}
(1)_{n}^\phi \\ (2)_{n}^\phi  
\end{gathered} \right]^n 
\overset{2}{\to}\left. \begin{gathered}
\left.\left.\begin{gathered}
(1)_{n\phi}\\(1')_{n\phi}  
\end{gathered}\right\}_{n}\right]^\phi \\ \left.\left.\begin{gathered}
(2)_{n\phi}\\ (2')_{n\phi}  
\end{gathered}\right\}_{n}\right]^\phi   
\end{gathered} \right]^n
\overset{3}{\to} \left.\begin{gathered}
(0)_{\phi}^n \\ (0')_{\phi}^n
\end{gathered}\right]^{\phi}
\overset{\phi|a}{=}
\left.\begin{gathered}
(0)_{a\uparrow}^n \\ (0') _{a}^n
\end{gathered}\right]^{a}
\overset{4}{\to}
\begin{gathered}
(0)^{na}\\ (0')^{na} 
\end{gathered}
\overset{a|b}{=} \begin{gathered}
(0) ^{na}\\ (0') ^{n}_{b\uparrow}
\end{gathered}
\overset{5}{\to} \begin{gathered}
(0) ^{na}\\ (0')^{nb}
\end{gathered}
\end{eqnarray}
\end{widetext}

The diagram has the following stages:
\begin{enumerate}
    \item The initial photon $(0)$ with a defined number of particles $n$ (we have exactly one photon) and a defined phase $\phi$ is split into two trajectories with qils $(1)$ and $(2)$. They have the same phase; however, each of them have their number of particle undefined with the sum of one (a constraint on degrees of freedom of $n$). That means that if we choose a trajectory (thus choosing a first or second qil), the number of particles on the other trajectory becomes $0$;
    \item Every qil splits into a signal and an idler parts. Their number of particles are \textit{chained} (the signal and the idler parts exist or are absent simultaneously) and their phases are \textit{entangled} (the phase sum is constant);
    \item We join the signal parts between each other, defining the number of particles to be $1$. Also we do the same with idlers. Thus we create a pair of qils with their phases entangled. Then we change the variables from $\phi$ to the eigenbasis of the $a$-detectors. This transform converts a continuous degree of freedom to a discrete one and is irreversible;
    \item The $a$-detectors click, raising the index of a signal qil (and raising the idler qil's index as well after the reduction). Then we change the basis of the idler to $b$-detectors' eigenbasis, obtaining the undefined $b$. In the bra-ket notation it means $\ket{0}_a \to \frac{1}{2}\left(\ket{0}_b + \sqrt{2}\ket{1}_b + \ket{3}_b\right)$ and $\ket{1}_a \to \frac{1}{2}\left(\ket{0}_b + \sqrt{2}\ket{2}_b + \ket{3}_b\right)$;
    \item Finally, we measure the idler qil, obtaining the desired statistics.
\end{enumerate}

It's remarkable that the states after the first and the third steps have the same structure with the indices $\phi$ and $n$ interchanged.

\subsection{PMT (qils of chained modes, measurement)}
\label{ch:PMT}

A photomultiplier tube (PMT) is a well-known detector of photons. The photon strikes out the first electron, which then accelerates to the first dynode and strikes out several other electrons. The process repeats leading to the significant growth of the number of electrons in movement. Since a PMT is a detector, we expect the event probability to be around $1$. Let's show that in our theory it is so.

\begin{figure*}
    \centering
    \includegraphics[width=0.7\linewidth]{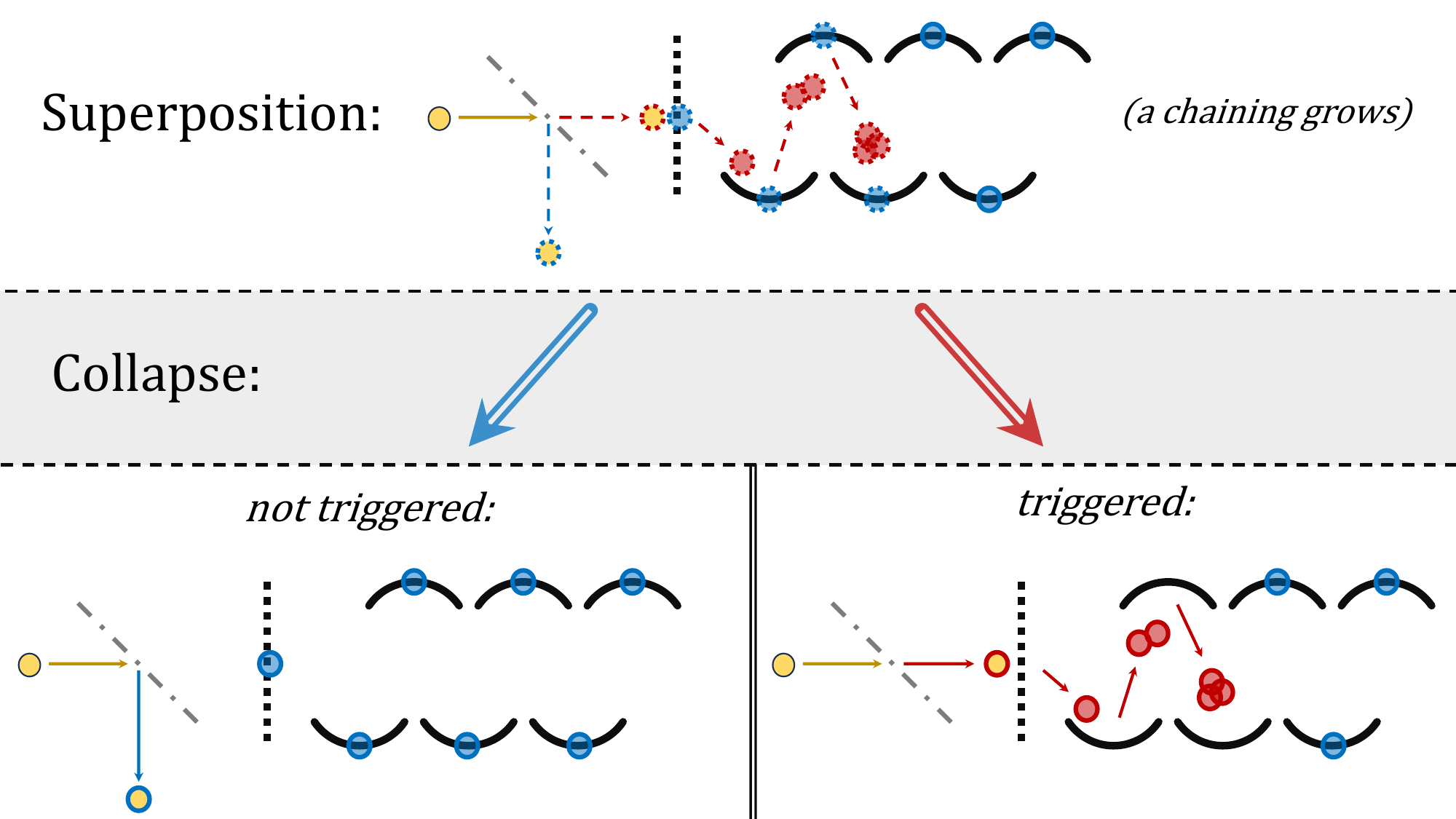}
    \caption{A PMT scheme. At first, we have a photon (yellow dot) that splits into a superposition of red and blue paths. The electrons then also split into two paths - red (the photon is detected) and blue (the photon is deflected by a mirror). Every path contains a set of chained electrons which grows without its number of degrees of freedom growing. At some moment, a collapse event occurs, and the chained degree of freedom falls into one of two possible options.}
    \label{fig:pmt}
\end{figure*}

As a model of a PMT we will consider a set of electrons:

\begin{eqnarray}
    \mathcal A = \begin{gathered}
        (e)^n\\ (e)^n\\ (e)^n\\ \dots
    \end{gathered}
\end{eqnarray}

Here the parameter $n$ represents the state of an electron (triggered/not triggered), which is also the occupation number of a free (triggered) electron mode. Initially all the electrons are in the state ''not triggered'' with $n = 0$. However, if one of them is triggered, it triggers the next with a $100\%$ efficiency. The first one is to be triggered by the initial photon.

Let's consider then a translucent mirror before a PMT which prepares the photon with the following qil-state:

\begin{eqnarray}
    (\gamma)^n \to \left.\begin{gathered}
        (r)_n\\ (t)_n
    \end{gathered}\right]^n
\end{eqnarray}

There're two parts: the transmitted photon $(t)$ (that goes to a PMT) and the reflected photon $(r)$ with the sum of their numbers of particles equal to $1$ (if one exists, the other vanishes).

The transmitted qil $(t)$ then gets into a PMT and triggers an avalanche of chainings:

\begin{eqnarray}
    \begin{gathered}
        \left.\begin{gathered}
        (r)_n\\ (t)_n
    \end{gathered}\right]^n \\ \begin{gathered}
        (e)^n\\ (e)^n\\ (e)^n\\ \dots
    \end{gathered}
    \end{gathered} \to \begin{gathered}
        \left.\begin{gathered}
        (r)_n\\ \left.\begin{gathered}
            (t)_n\\ (e)_n
        \end{gathered}\right\}_n
    \end{gathered}\right]^n \\ \begin{gathered}
        (e)^n\\ (e)^n\\ \dots
    \end{gathered}
    \end{gathered} \to \begin{gathered}
        \left.\begin{gathered}
        (r)_n\\ \left.\begin{gathered}
            (t)_n\\(e)_n\\ (e)_n
        \end{gathered}\right\}_n
    \end{gathered}\right]^n \\ \begin{gathered}
         (e)^n\\ \dots
    \end{gathered}
    \end{gathered} \to \dots
\end{eqnarray}

This diagram has the following meaning. The reflected photon mode with the $(r)$-qil doesn't affect the electrons. However, the other mode with the qil $(t)$ chains with the first electron along $n$, which means that if $(t)$-mode is occupied, the electron is triggered, and vice versa. The first electron chains with the second, and so on. 

If the PMT has the efficiency about $N = 10^6$, the probability of the event happening an \textit{some stage} of this process is $P = 1 - \left(1 - \frac{1}{\Sigma}\right)^N$. Even if $\Sigma \gg 1$, $P\approx 1$. It's also easy to show that this scheme leads to an almost guaranteed event even without a Chinese Whispers rule:

\begin{eqnarray}
    P(\text{event}) = 1 - \prod_i\left(1 - P_i\right) \approx 1,
\end{eqnarray}

where $P_i$ is a probability of an event at the appearance of an $i$-th chaining.

\begin{figure*}
    \centering
    \includegraphics[width=0.7\linewidth]{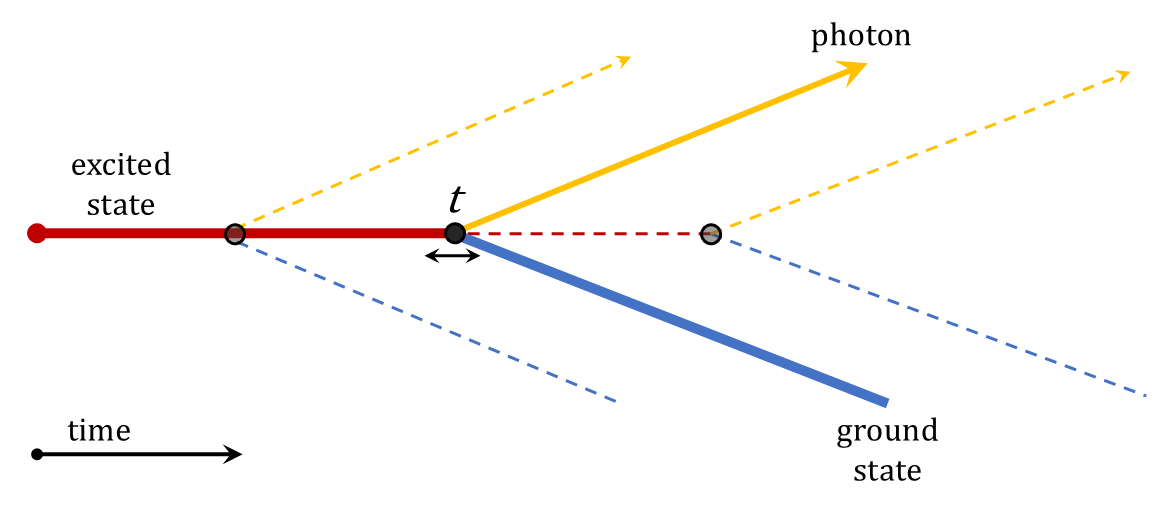}
    \caption{An illustration of a time-state corresponding the atomic relaxation process. The excited atom's world line (red) is in a superposition of states with different ending time point $t$. The ground state's world line (blue) and the one of a photon (yellow) are in a superposition as well. However, their initial time points $t$ are both correspond to the ending time point of the excited state, which means the world lines' qils are chained.}
    \label{fig:ziptime}
\end{figure*}

\subsection{Atomic relaxation (qils of time-states, non-measurement event)}

\label{ch:decay}

Let's consider now an excited atom, which is able to relax into its ground state with emission of a photon. The qil-diagram of this process connects qils of world lines of the excited $(*)$ and ground $(\cdot)$ states of the atom and the photon $(\gamma)$ with a moment of decay $t$ (see Fig. \ref{fig:ziptime}). Note that now we consider qils not for the systems' \textit{states evolving in time}, but for the systems' whole world lines. Thus we consider \textit{time-states} instead of usual wave-functions to be objects for qils.

The unitary decay process starts from the moment the excited atom appears and lasts for the infinity. The probability distribution of the excited atom at the moment $t$ is

\begin{eqnarray}
    P = e^{-t/\tau}/\tau,
\end{eqnarray}

where $\tau$ is the excited state's lifetime. The qil-diagram for the full time-state is

\begin{eqnarray}
    \begin{gathered}
        \left.\begin{gathered}
        (*)_t\\
        (\cdot)_t\\
        (\gamma)_t
    \end{gathered}\right\}_t
    \end{gathered}
\end{eqnarray}

Since it is a chaining qil, it has a $1/\Sigma$ probability to spawn with $t$ defined. Thus, such relaxation process also leads to non-unitarity in some cases.

\subsection{A cat and a mirror (chaining vs. entanglement)}

Let's consider a classical setup (''Schrödinger's cat''). A cat, a flask of poison and a Geiger counter are put in a box with a radioactive atom. After $\tau$ seconds the atom gets into an even superposition of $\ket{\cdot}$ (decayed state) and  $\ket{*}$ (instable state). In an original formulation given by Schrödinger \cite{Schrodinger1935}, the Geiger counter also gets into a superposition (triggered/not triggered). The triggered Geiger counter opens a poison, and the cat dies. Thus, the cat also gets into a superposition of living and dead states.

Let's write a qil-diagram of the experiment, analogously to a subsection \ref{ch:PMT}. The trigger of a Geiger counter is a process of lots of successive acts of ionization, where a previous triggers the next. Thus, the Geiger counter has the following qil-diagram:

\begin{widetext}
\begin{eqnarray}
    \begin{gathered}
        \left.\begin{gathered}
            (*)_n\\(\cdot)_n
        \end{gathered}\right]^n\\ (+)^n\\(+)^n\\\dots 
    \end{gathered}\rightarrow \begin{gathered}
        \left.\begin{gathered}
            (*)_n\\\left.\begin{gathered}
                (\cdot)_n\\ (+)_n
            \end{gathered}\right\}_n
        \end{gathered}\right]^n\\(+)^n\\\dots 
    \end{gathered} \rightarrow\begin{gathered}
        \left.\begin{gathered}
            (*)_n\\\left.\begin{gathered}
                (\cdot)_n\\ (+)_n\\(+)_n
            \end{gathered}\right\}_n
        \end{gathered}\right]^n\\\dots 
    \end{gathered}\rightarrow \dots\rightarrow \left.\begin{gathered}
        (*)_n\\\left.\begin{gathered}
            (\cdot)_n\\ (G)_n
        \end{gathered}\right\}_n
    \end{gathered}\right]^n
\end{eqnarray}
\end{widetext}

where $(+)$ is a qil of an ion with an existence number $n$ ($1$ is for existing, $0$ is for absence) and $G$ is a qil of a Geiger counter's signal.

For the cat to be included to the system, one must divide the cat's body into a number of subsystems since the cat doesn't interact with the poison immediately. At first, the poison gets into the cat's lungs, changing their state; then, it gets into the blood, reaches the nervous system; finally, by destroying the nervous system, the poison stops the heart, and so on. On each step the poison chains with lots and lots of quantum particles of which the subsystem consists. So, if we consider the qils for every poison-affected subsystem ($(L)$, $(B)$ and so on for the lungs, the blood and others), we can write in general

\begin{widetext}
\begin{eqnarray}
    \begin{gathered}
        \left.\begin{gathered}
        (*)_n\\\left.\begin{gathered}
            (\cdot)_n\\ (G)_n
        \end{gathered}\right\}_n
    \end{gathered}\right]^n\\(L)^n\\(B)^n\\\dots
    \end{gathered} \rightarrow\dots\rightarrow\begin{gathered}
        \left.\begin{gathered}
        (*)_n\\\left.\begin{gathered}
            (\cdot)_n\\ (G)_n\\(L)_n
        \end{gathered}\right\}_n
    \end{gathered}\right]^n\\(B)^n\\\dots
    \end{gathered}\rightarrow\dots\rightarrow\begin{gathered}
        \left.\begin{gathered}
        (*)_n\\\left.\begin{gathered}
            (\cdot)_n\\ (G)_n\\(L)_n\\(B)_n
        \end{gathered}\right\}_n
    \end{gathered}\right]^n\\\dots
    \end{gathered}\rightarrow\dots\rightarrow
        \left.\begin{gathered}
        (*)_n\\\left.\begin{gathered}
            (\cdot)_n\\ (G)_n\\(\dagger)_n
        \end{gathered}\right\}_n
    \end{gathered}\right]^n
\end{eqnarray}
\end{widetext}

where $(\dagger)$ is a qil of a dead cat.

The overall diagram expose the following: the Schrödinger's cat experiments contains lots and lots of chainings. Even with a large $\Sigma$ that number of chainings guarantees an event, thus making a cat in superposition of its living and dead states nearly impossible. In the other hand, such macroscopic objects as mirrors don't trigger events at all, since the process of reflection doesn't cause any chainings. In the case of inelastic reflection the overall interaction of a particle with a qil $(\psi)$ on a reflector $(M)$ reduces simply to an entanglement of momenta:

\begin{eqnarray}
    \begin{gathered}
        (\psi)^p\\(M)^p
    \end{gathered}\rightarrow \left.\begin{gathered}
        (\psi)_p\\(M)_p
    \end{gathered}\right]^p
\end{eqnarray}

In other words, to distinguish a cat from a mirror, one must decide if the system has large number of chainings as a response to a quantum interaction, or not.

\section{Experimental estimation for $\Sigma$}
\label{ch:experimental}

Let now consider a compound particle interacting with an ideal beam splitter that has a probability of deflecting a particle out of the initial direction. Then this beam splitter triggers a chaining:

\begin{eqnarray}
    \begin{gathered}
        (1)^n\\(2)^n \\(3)^n
    \end{gathered} \to \left.\begin{gathered}
        \left.\begin{gathered}
            (1|a)_n\\ (2|a)_n \\ (3|a)_n
        \end{gathered}\right\}_n\\ \left.\begin{gathered}
            (1|b)_n\\ (2|b)_n \\ (3|b)_n
        \end{gathered}\right\}_n
    \end{gathered}\right]^n
\end{eqnarray}

where the letter $a$ denotes the passed particle and the $b$ is for the deflected one. Since the particle is solid, there's no possibility for a part of their components to be deflected while the rest is passed. So there's a probability $P = \frac{1}{\Sigma}$ for a path superposition to break, and thus making the interference impossible. If we then construct a Mach - Zehnder interferometer \cite{zehnder} with such a beam splitter, the particles affected by such a break will not form the interference pattern, forming a bias with the integral proportional to $P$. If we denote a bias integral as $B \sim P$, interference integral as $Q \sim 1-P$ and the total integral (interference plus bias) as $I \sim 1$, we may extract the $\Sigma$ out of the experimental data:

\begin{eqnarray}
    \Sigma = \frac{1}{P} = \frac{Q+B}{B} = \frac{I}{B}.
\end{eqnarray}

The same is true for a double-slit experiment with a diagram

\begin{eqnarray}
    \begin{gathered}
        (1)^n\\(2)^n \\(3)^n
    \end{gathered} \to \left.\begin{gathered}
        \left.\begin{gathered}
            (1|s)_n\\ (2|s)_n \\ (3|s)_n
        \end{gathered}\right\}_n\\ \left.\begin{gathered}
            (1|l)_n\\ (2|l)_n \\ (3|l)_n
        \end{gathered}\right\}_n\\ \left.\begin{gathered}
            (1|r)_n\\ (2|r)_n \\ (3|r)_n
        \end{gathered}\right\}_n
    \end{gathered}\right]^n
\end{eqnarray}

where $s$-particles are stuck on a screen and $l$- and $r$-particles are the particles going to the left or the right slit correspondingly. Here we may also extract $\Sigma$ as a ratio of a total integral $I$ and the integral of the bias $B$ as $\Sigma = \frac{I}{B}$.

This \textit{collapse effect} must occur in any interference experiment conducted with compound particles. One may note that its intensity does not depend on the particles' content and is as relevant for neutrons as for atoms and molecules. However, the collapse effect may mix with a well-known \textit{decoherence} since the latter induces qualitatively the same bias. Thus, the $\Sigma$ extimated from the experiment may be \textit{understated}.

Such experiments were conducted with:
\begin{itemize}
    \item neutrons on a beam splitter \cite{n1s} ($\Sigma \geq 1.28 \pm 0.07$);
    \item neutrons on a double-slit \cite{n2s} ($\Sigma \geq 1.54\pm 0.04$);
    \item Helium atoms on a double-slit \cite{He} ($\Sigma \geq 1.21\pm 0.29$);
    \item Neon atoms on a double-slit \cite{Ne} ($\Sigma \geq 1.4\pm 0.13$).
\end{itemize}

\begin{figure*}
    \begin{center}
	\begin{minipage}{0.45\linewidth}
  		\includegraphics[width=1\linewidth]{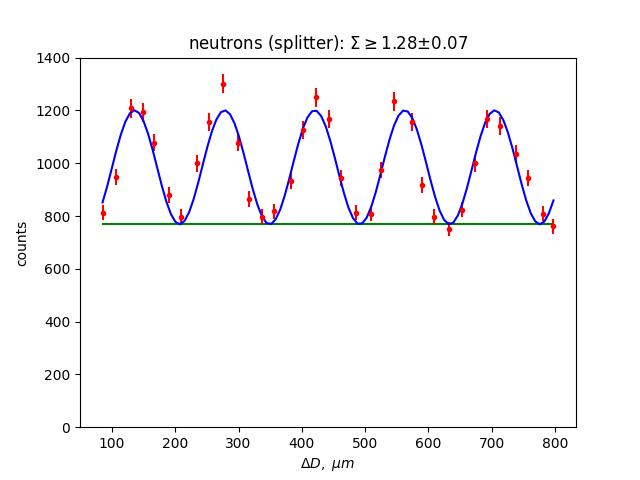}
   	\end{minipage}
   	\hfill
   	\begin{minipage}{0.45\linewidth}
   		\includegraphics[width=1\linewidth]{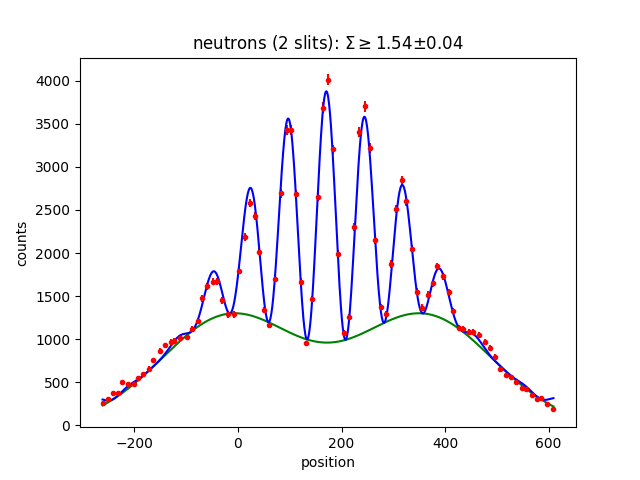}
   	\end{minipage}
        \hfill
        \begin{minipage}{0.45\linewidth}
  		\includegraphics[width=1\linewidth]{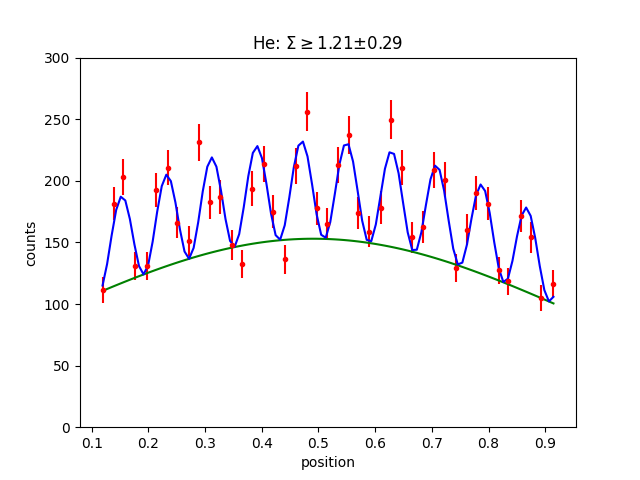}
   	\end{minipage}
   	\hfill
   	\begin{minipage}{0.45\linewidth}
   		\includegraphics[width=1\linewidth]{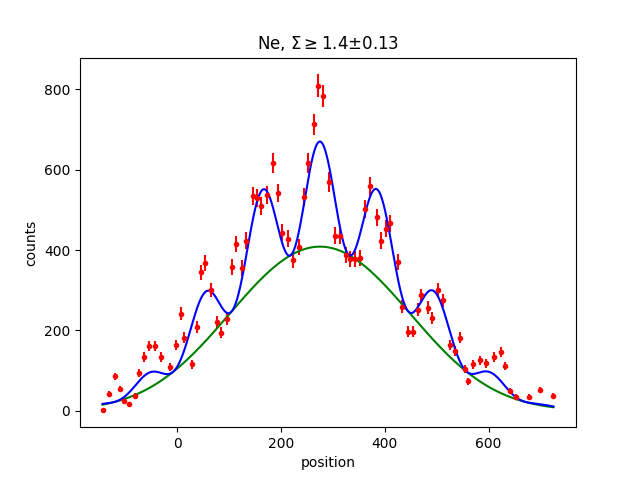}
   	\end{minipage}
    \end{center}
    \caption{Fits for the interference patterns of several experiments. The blue line is a general fit, the green line is the bias evaluated and the red dots are the experimental data.}
    \label{fig:expts}
\end{figure*}

The fits for the experiments' interference patterns are shown on Fig. \ref{fig:expts}. Considering these results and a possibility of understating one may assume the $\Sigma$ to have a value about $1.5$.

\begin{table*}
    \centering
    \begin{tabular}{|p{0.9cm}||p{2.3cm}|p{2.3cm}|p{2.3cm}|p{2.3cm}|p{2.3cm}|p{3cm}|}
    \hline
        Setup & $\text{n}$, beam splitter & $\text{n}$, double-slit & $\text{He}$, double-slit & $\text{Ne}$, double-slit & Delayed Choice QE \\
    \hline\hline
        $\Sigma_\text{est.}$ & $\geq1.28\pm 0.07$ & $\geq 1.54\pm 0.04$ & $\geq 1.21\pm 0.29$ & $\geq 1.4\pm 0.13$ & $\geq 1.26\pm 0.34$ \\
    \hline
    \end{tabular}
    \caption{the lower boundaries of $\Sigma$ extimated by different experiments. Note that in the case of significant decoherence the non-interfering bias may be higher, thus making the value for $\Sigma$ lower.}
\end{table*}

Finally, one may recollect the diagram describing the Delayed-Choice Quantum Eraser experiment:

\begin{widetext}
\begin{eqnarray}
    (0)^{n\phi} 
\overset{1}{\to} \left. \begin{gathered}
(1)_{n}^\phi \\ (2)_{n}^\phi  
\end{gathered} \right]^n 
\overset{2}{\to}\left. \begin{gathered}
\left.\left.\begin{gathered}
(1)_{n\phi}\\(1')_{n\phi}  
\end{gathered}\right\}_{n}\right]^\phi \\ \left.\left.\begin{gathered}
(2)_{n\phi}\\ (2')_{n\phi}  
\end{gathered}\right\}_{n}\right]^\phi   
\end{gathered} \right]^n
\overset{3}{\to} \left.\begin{gathered}
(0)_{\phi}^n \\ (0')_{\phi}^n
\end{gathered}\right]^{\phi}
\overset{\phi|a}{=}
\left.\begin{gathered}
(0)_{a\uparrow}^n \\ (0') _{a}^n
\end{gathered}\right]^{a}
\overset{4}{\to}
\begin{gathered}
(0)^{na}\\ (0')^{na} 
\end{gathered}
\overset{a|b}{=} \begin{gathered}
(0) ^{na}\\ (0') ^{n}_{b\uparrow}
\end{gathered}
\overset{5}{\to} \begin{gathered}
(0) ^{na}\\ (0')^{nb}
\end{gathered}
\end{eqnarray}
\end{widetext}

One should notice that the step $2$ is, in fact, \textit{two distinct transitions}. Unlike the beam splitter and the double-slit diagrams, these two chainings appear independently, thus triggering at least one event with probability of

\begin{eqnarray}
    P = 1 - \left(1 - \frac{1}{\Sigma}\right)^2.
\end{eqnarray}

As in the previous examples, either of these two events breaks the interference pattern. For example we consider a case where the second trajectory totally dies out:

\begin{eqnarray}
    (0)^{n\phi} 
\overset{1}{\to} \left. \begin{gathered}
(1)_{n}^\phi \\ (2)_{n}^\phi  
\end{gathered} \right]^n 
\overset{2}{\to}\left. \begin{gathered}
\left.\left.\begin{gathered}
(1)_{n\phi}\\(1')_{n\phi}  
\end{gathered}\right\}_{n\uparrow}\right]^\phi \\ \left.\left.\begin{gathered}
(2)_{n\phi}\\ (2')_{n\phi}  
\end{gathered}\right\}_{n}\right]^\phi   
\end{gathered} \right]^n = 
\left.\begin{gathered}
    (1)_{\phi}^n\\ (1')_\phi^n
\end{gathered}\right]^\phi
\end{eqnarray}

Unlike the original case, this resulting qil has its signal subqil preset in a state with a $0.5:0.5$ pattern on the detector $a$. Simply calculating the statistics on the detector $b$ for any outcome on the $a$ one may obtain that for the interference pattern on $b1$ the probability of an event is

\begin{eqnarray}
    P = \frac{2B}{I}.
\end{eqnarray}

\begin{figure}
    \centering
    \includegraphics[width=1\linewidth]{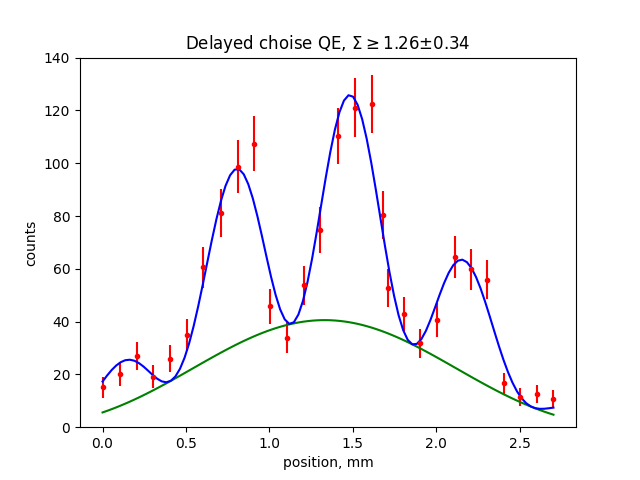}
    \caption{Fits for the interference pattern of a Delayed Choice Quantum Eraser experiment \cite{dcqe}. The blue line is a general fit, the green line is the bias evaluated and the red dots are the experimental data.}
    \label{fig:DCQEint}
\end{figure}

The fit for the interference pattern from the original article about the experiment \cite{dcqe} is depicted on Fig. \ref{fig:DCQEint}. The value of $\Sigma \geq 1.26\pm 0.34$ also agrees with the estimation of $\Sigma \approx 1.5$.

\textbf{N. B.} It's still uncertain if $\Sigma$ has the value of $1.5$ or not. The future experiments may evaluate the higher value of $\Sigma$, if they'll manage to further suppress the proper decoherence factors of the experiments. Moreover, the method presented is only capable to evaluate the lower bound for $\Sigma$.

\section{Discussion}
\label{ch:discussion}

The model proposed has several strong points. At first, it suggests not an interpretation of quantum mechanics, but a new objective collapse theory. Its three theses allow to solve the measurement problem not by drawing a scale threshold of the quantum world, but considering a number of acts of chainings needed to construct a classical system. Thus, the classicality is a property not of size, but of a process of \textit{reaching} that size expressed in terms of chainings.

This point radically distinguishes our theory from the classical examples such as CSL or DP models. These models predict that the loss of coherence somehow depends or the generalized ''size'' of a system; however, the examples given in the section \ref{ch:experimental} show that the non-coherent bias has the same order for systems of radically different ''sizes'' (from neutrons to neon atoms). Although it can still be explained by fine-tuning of the experiments' parameters, our model presents a more universal and significantly simpler explanation. This explanation showed a good correspondence with experimental data, encompassing not only interference experiments, but a Delayed-choice quantum eraser as well. The theory present makes this generalization not only possible, but also convenient since it provides us with qil-diagrams, the main purpose of which is to account interconnections between systems' degrees of freedom.

The theory considered in this article still has a room for improvement. Although it manages to properly describe atomic relaxation (see Section \ref{ch:decay}), it's necessary to develop a general method to operate continuous systems and processes (such as quantum fields). However, such an improvement is out of the scope of this paper and would be considered in our subsequent works. It's also possible that the value of $\Sigma$ depends on some parameters of the experiment considered, making the ChiWhi rule just an approximation of the real physical law. Finally, the $\Sigma$ estimation method proposed, in fact, gives only a lower bound on its value, relying only on the hope that we indeed deal with the clearest experimental data obtained. If that hope is not unfounded, the true value of $\Sigma$ is about $1.5$; otherwise, it would be greater. In fact, its value can be arbitrarily large.

The result obtained in this article hints on a surprising observation. Any compound system, being put in an experimental setup that splits it along a new degree of freedom $p$, has its components chained by an occupation number:

\begin{eqnarray}
    \begin{gathered}
        (1)\\ \dots\\ (m)
    \end{gathered} \rightarrow \left.\begin{gathered}
        \left.\begin{gathered}
            (1)^p_n\\ \dots\\(m)^p_n
        \end{gathered}\right\}_n\\ \dots\\ \left.\begin{gathered}
            (1)^p_n\\ \dots\\(m)^p_n
        \end{gathered}\right\}_n
    \end{gathered}\right]^n
\end{eqnarray}

where each chaining corresponds to a different value of $p$. It leads to an excessive non-unitarity for most such systems. Consequently, if a quantum computer uses elementary particles as its qubits, its coherence time could be significantly higher, as one source of destabilization — chaining-induced collapse — would be absent.

\section{Acknowledgments}

R. V. L. (Voevodsky Institute of Chemical Kinetics and Combustion SB RAS) acknowledge the core funding from the Russian Federal Ministry of Science and Higher Education (FWGF-2026-0007, 126020516710-7).

\bibliographystyle{unsrt}
\bibliography{bib}

@article{dcqe,
  doi = {10.48550/ARXIV.QUANT-PH/9903047},
  url = {https://arxiv.org/abs/quant-ph/9903047},
  author = {Kim,  Yoon-Ho and Yu,  R. and Kulik,  S. P. and Shih,  Y. H. and Scully,  Marlan . O.},
  title = {A Delayed Choice Quantum Eraser},
  publisher = {arXiv},
  year = {1999},
  copyright = {Assumed arXiv.org perpetual,  non-exclusive license to distribute this article for submissions made before January 2004}
}

@article{Frantsuzov2020,
  author = {P. A. Frantsuzov, A. V. Nechiporenko},
  ISSN = {1560-7488},
  url = {http://dx.doi.org/10.15372/PS20200207},
  DOI = {10.15372/ps20200207},
  number = {2},
  journal = {Философия науки},
  publisher = {Publishing House SB RAS},
  year = {2020}
}

@article{zehnder,
    author = {Zehnder, L},
    title = {Ein neuer Interferenzrefraktor},
    journal = {Zeitschrift für Instrumentenkunde},
    year = {1891}
}

@article{n1s,
title = {Test of a single crystal neutron interferometer},
journal = {Physics Letters A},
volume = {47},
number = {5},
pages = {369-371},
year = {1974},
issn = {0375-9601},
doi = {https://doi.org/10.1016/0375-9601(74)90132-7},
url = {https://www.sciencedirect.com/science/article/pii/0375960174901327},
author = {H. Rauch and W. Treimer and U. Bonse}
}

@article{n2s,
  title = {Single- and double-slit diffraction of neutrons},
  volume = {60},
  ISSN = {0034-6861},
  url = {http://dx.doi.org/10.1103/RevModPhys.60.1067},
  DOI = {10.1103/revmodphys.60.1067},
  number = {4},
  journal = {Reviews of Modern Physics},
  publisher = {American Physical Society (APS)},
  author = {Zeilinger,  Anton and G\"{a}hler,  Roland and Shull,  C. G. and Treimer,  Wolfgang and Mampe,  Walter},
  year = {1988},
  month = oct,
  pages = {1067–1073}
}

@article{He,
  title = {Young's double-slit experiment with atoms: A simple atom interferometer},
  author = {Carnal, O. and Mlynek, J.},
  journal = {Phys. Rev. Lett.},
  volume = {66},
  issue = {21},
  pages = {2689--2692},
  numpages = {0},
  year = {1991},
  month = {May},
  publisher = {American Physical Society},
  doi = {10.1103/PhysRevLett.66.2689},
  url = {https://link.aps.org/doi/10.1103/PhysRevLett.66.2689}
}

@article{Ne,
  title = {Double-slit interference with ultracold metastable neon atoms},
  volume = {46},
  ISSN = {1094-1622},
  url = {http://dx.doi.org/10.1103/PhysRevA.46.R17},
  DOI = {10.1103/physreva.46.r17},
  number = {1},
  journal = {Physical Review A},
  publisher = {American Physical Society (APS)},
  author = {Shimizu,  Fujio and Shimizu,  Kazuko and Takuma,  Hiroshi},
  year = {1992},
  month = jul,
  pages = {R17–R20}
}

@article{Einstein1935,
  title = {Can Quantum-Mechanical Description of Physical Reality Be Considered Complete?},
  volume = {47},
  ISSN = {0031-899X},
  url = {http://dx.doi.org/10.1103/PhysRev.47.777},
  DOI = {10.1103/physrev.47.777},
  number = {10},
  journal = {Physical Review},
  publisher = {American Physical Society (APS)},
  author = {Einstein,  A. and Podolsky,  B. and Rosen,  N.},
  year = {1935},
  month = may,
  pages = {777–780}
}

@article{Schrodinger1935,
  title = {Die gegenwärtige Situation in der Quantenmechanik},
  volume = {23},
  ISSN = {1432-1904},
  url = {http://dx.doi.org/10.1007/BF01491891},
  DOI = {10.1007/bf01491891},
  number = {48},
  journal = {Die Naturwissenschaften},
  publisher = {Springer Science and Business Media LLC},
  author = {Schrödinger,  E.},
  year = {1935},
  month = nov,
  pages = {807–812}
}

@book{Baggott:2011zz,
    author = "Baggott, Jim",
    title = "{The quantum story: A history in 40 moments}",
    year = "2011",
    publisher = "Oxford University Press"
}

@article{Wheeler2017,
  title = {Tales of the Quantum: Understanding Physics’ Most Fundamental Theory},
  volume = {85},
  ISSN = {1943-2909},
  url = {http://dx.doi.org/10.1119/1.4996872},
  DOI = {10.1119/1.4996872},
  number = {9},
  journal = {American Journal of Physics},
  publisher = {American Association of Physics Teachers (AAPT)},
  author = {Wheeler,  Nicholas},
  year = {2017},
  month = sep,
  pages = {722–723}
}

@book{OMNS2020,
  title = {Understanding Quantum Mechanics},
  ISBN = {9780691004358},
  url = {http://dx.doi.org/10.2307/j.ctv173f2pm},
  DOI = {10.2307/j.ctv173f2pm},
  publisher = {Princeton University Press},
  author = {OMNÈS,  ROLAND},
  year = {2020},
  month = dec 
}

@article{Zurek2003,
  title = {Decoherence,  einselection,  and the quantum origins of the classical},
  volume = {75},
  ISSN = {1539-0756},
  url = {http://dx.doi.org/10.1103/revmodphys.75.715},
  DOI = {10.1103/revmodphys.75.715},
  number = {3},
  journal = {Reviews of Modern Physics},
  publisher = {American Physical Society (APS)},
  author = {Zurek,  Wojciech Hubert},
  year = {2003},
  month = may,
  pages = {715–775}
}

@misc{zurek2003decoherencetransitionquantumclassical,
      title={Decoherence and the transition from quantum to classical -- REVISITED}, 
      author={Wojciech H. Zurek},
      year={2003},
      eprint={quant-ph/0306072},
      archivePrefix={arXiv},
      primaryClass={quant-ph},
      url={https://arxiv.org/abs/quant-ph/0306072}, 
}

@book{Chalmers1996-CHATCM-18,
	address = {New York},
	author = {David John Chalmers},
	editor = {},
	title = {The Conscious Mind: In Search of a Fundamental Theory},
	year = {1996}
}

@book{Kirk2005-KIRZAC-2,
	address = {Oxford, GB},
	author = {Robert Kirk},
	editor = {},
	publisher = {Oxford University Press UK},
	title = {Zombies and Consciousness},
	year = {2005}
}

@book{Penrose1989,
author = {Penrose, Roger},
title = {The emperor's new mind: concerning computers, minds, and the laws of physics},
year = {1989},
isbn = {0198519737},
publisher = {Oxford University Press, Inc.},
address = {USA}
}

@misc{schreiber1995livesschroedingerscat,
      title={The Nine Lives of Schroedinger's Cat}, 
      author={Zvi Schreiber},
      year={1995},
      eprint={quant-ph/9501014},
      archivePrefix={arXiv},
      primaryClass={quant-ph},
      url={https://arxiv.org/abs/quant-ph/9501014}, 
}

@article{Okon_2014,
   title={Benefits of Objective Collapse Models for Cosmology and Quantum Gravity},
   volume={44},
   ISSN={1572-9516},
   url={http://dx.doi.org/10.1007/s10701-014-9772-6},
   DOI={10.1007/s10701-014-9772-6},
   number={2},
   journal={Foundations of Physics},
   publisher={Springer Science and Business Media LLC},
   author={Okon, Elias and Sudarsky, Daniel},
   year={2014},
   month=feb, pages={114–143} }

@article{Disi1989,
  title = {Models for universal reduction of macroscopic quantum fluctuations},
  volume = {40},
  ISSN = {0556-2791},
  url = {http://dx.doi.org/10.1103/PhysRevA.40.1165},
  DOI = {10.1103/physreva.40.1165},
  number = {3},
  journal = {Physical Review A},
  publisher = {American Physical Society (APS)},
  author = {Diósi,  L.},
  year = {1989},
  month = aug,
  pages = {1165–1174}
}

@article{Penrose1996,
  title = {On Gravity’s role in Quantum State Reduction},
  volume = {28},
  ISSN = {1572-9532},
  url = {http://dx.doi.org/10.1007/BF02105068},
  DOI = {10.1007/bf02105068},
  number = {5},
  journal = {General Relativity and Gravitation},
  publisher = {Springer Science and Business Media LLC},
  author = {Penrose,  Roger},
  year = {1996},
  month = may,
  pages = {581–600}
}

@article{Pearle1989,
  title = {Combining stochastic dynamical state-vector reduction with spontaneous localization},
  volume = {39},
  ISSN = {0556-2791},
  url = {http://dx.doi.org/10.1103/PhysRevA.39.2277},
  DOI = {10.1103/physreva.39.2277},
  number = {5},
  journal = {Physical Review A},
  publisher = {American Physical Society (APS)},
  author = {Pearle,  Philip},
  year = {1989},
  month = mar,
  pages = {2277–2289}
}

@article{Ghirardi1990,
  title = {Markov processes in Hilbert space and continuous spontaneous localization of systems of identical particles},
  volume = {42},
  ISSN = {1094-1622},
  url = {http://dx.doi.org/10.1103/PhysRevA.42.78},
  DOI = {10.1103/physreva.42.78},
  number = {1},
  journal = {Physical Review A},
  publisher = {American Physical Society (APS)},
  author = {Ghirardi,  Gian Carlo and Pearle,  Philip and Rimini,  Alberto},
  year = {1990},
  month = jul,
  pages = {78–89}
}

\end{document}